\documentclass[a4paper]{mem}
\usepackage{natbib}
\usepackage{graphicx}
\usepackage[a4paper]{hyperref}
\idline{74}{ 1}
\begin{document}
\title{Color Variability of AGNs}
\author{F. Vagnetti \inst{1}, \and D. Trevese \inst{2}}

\offprints{F. Vagnetti}

\institute{Dipartimento di Fisica, Universit\`a di Roma ``Tor Vergata'',
Via della Ricerca Scientifica 1, I-00133 Roma, Italy --
\email{fausto.vagnetti@roma2.infn.it}\\
               \and
Dipartimento di Fisica, Universit\`a di Roma ``La Sapienza'',
Piazzale A. Moro 2, I-00185 Roma, Italy --
\email{dario.trevese@roma1.infn.it}
}

\abstract{Optical spectral variability of quasars and BL Lac Objects is
compared by means of the spectral variability parameter $\beta$ \citep{tre02}.
Both kinds of objects change their spectral slopes $\alpha$, becoming bluer
when brighter, but BL Lac Objects have smaller $\beta$ values and are
clearly separated from quasars in the $\alpha-\beta$
plane. Models accounting for the origin of the variability are discussed for
both classes of objects.

    }
    \authorrunning{F. Vagnetti and D. Trevese}
    \titlerunning{Color Variability of AGNs}
    \maketitle
%

\section{Introduction}
Variability of the spectral energy distribution (SED) of
Active Galactic Nuclei (AGNs) is a powerful tool to investigate the
role of the main emission processes in different AGN
classes, and the origin of their variations.
The most common behavior in the optical band is that AGNs
become bluer, i.e. their spectrum becomes harder,
when brighter. This has been shown for individual Quasi Stellar Objects (QSOs)
and Seyferts \citep{cut85,ede90,kin91,pal94} and for one complete sample, i.e.
for the 42 PG QSOs monitored by \citet{giv99} in B and R for 7 
years. Evidence based on two epochs has been found also for the faint
QSO sample in the SA 57 \citep{tre01}. The same
trend is apparently shared by BL Lac Objects, as shown by \citet{dam02},
who present 5-year long B, V, R, I light curves for eight objects.
A quantitative estimate of color variability is necessary to compare the
observational data with emission models and to compare different AGN classes.
In a previous paper \citep{tre02}, we introduced the spectral variability
parameter $\beta \equiv \Delta \alpha / \Delta \log f_{\nu}$,
$f_{\nu}$ being the specific flux and
$\alpha\equiv\partial\log f_{\nu}/\partial\log\nu$ the spectral slope.
From the average $\alpha$ and $\beta$ values, it was possible to
derive constraints on the variability mechanisms.
The spectral variability parameter was then estimated \citep{vag02}
for the BL Lac Objects of \citet{dam02}. Observations and models can 
be compared in the $\alpha-\beta$ plane, which is reported in Fig. 1
for both the PG QSOs and the BL Lac Objects.

\section{Quasars}
Various models have been compared with the spectral variability of
PG QSOs \citep{tre02} and are shown in Fig. 1.

(i) The dot-dashed line represents the spectral variability
due to small temperature changes for a sequence of black bodies
with different temperatures (increasing from left to right).

(ii) We evaluate the effect of the host galaxy through numerical simulations
based on templates of the QSO and host galaxy SEDs,
derived from the atlas of normal QSO continuum spectra \citep{elv94}.
We added to the fixed host galaxy template SED the average QSO spectrum with a
relative weight measured by the parameter $\eta \equiv  \log({L_H^Q}/{L_H^g})$,
where ${L_H^Q}$ and ${L_H^g}$ are the total $H$ band luminosities of the  
QSO and the host galaxy respectively. 
Variability is represented by small changes $\Delta \eta$, around each
$\eta$ value, with an amplitude corresponding
to a r.m.s. variability $\sigma_B = 0.16$ mag in the blue band. 
The result is shown in Fig. 1 for $-3 < \eta < 3$ and for $z=0$, 
represented by a thin, continuous line, and clearly
shows that the effect of the host galaxy is not sufficient to account for
the observed changes of the spectral slope. 

(iii) We considered the accretion disk model of \citet{sie95}, 
corresponding to a Kerr metric and modified black body 
SED, which depend on the black hole mass $M$, the accretion rate $\dot{M}$ and
the inclination $\theta$.
A change of $\dot{m} \equiv \dot{M} c^2/L_E$ ($L_E$ being the Eddington luminosity)
produces a variation of both  
luminosity and the SED shape. The result is represented in Fig. 1 by a thick, dashed
curve, for $\dot{m}$ varying
between 0.1 and 0.3, and for $\theta=0$ (face on disk). 
The spectral variations are clearly smaller,
on average, than the observed ones. This means that a transition e.g. from
a lower to a higher $\dot{M}$ regime implies a larger luminosity change
for a given slope variation, respect to what is observed.

(iv) Transient phenomena, like hot spots produced
on the accretion disk by instability phenomena \citep{kaw98},
instead of a transition to a new equilibrium state,
may better explain the relatively large changes of the local spectral slope.
We use a simple model based on the  addition
of a black body flare to the disk SED, represented by the average 
QSO SED of \citet{elv94}. The result is shown in Fig. 1 by the large filled
squares, corresponding to hot spots with $T \approx 10^5$ K (upper),
and $T \approx 6\cdot 10^4$ K (lower).

    \begin{figure*}
    \centering
    \resizebox{\hsize}{!}{{\includegraphics{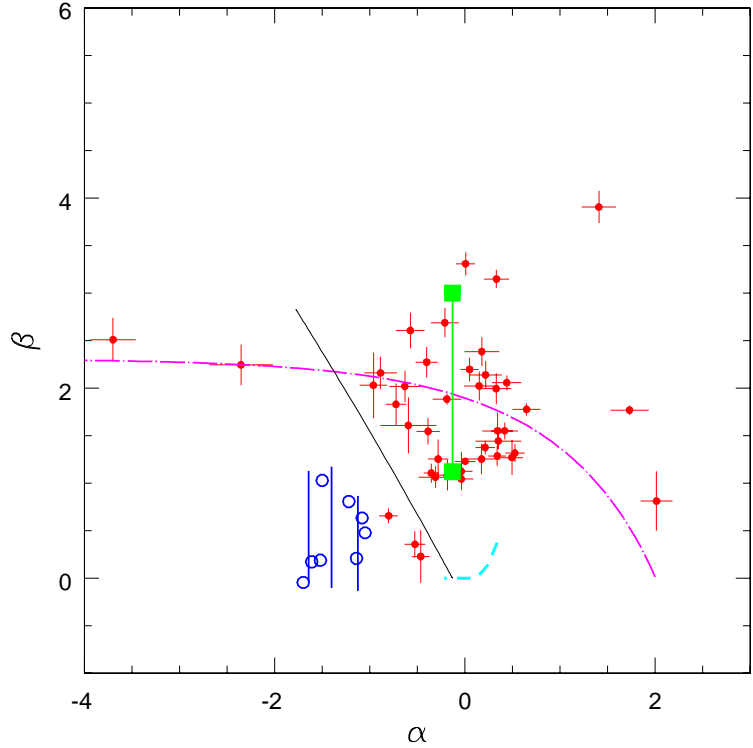}}}
    \caption{The spectral variability parameter $\beta$
versus the average spectral slope $\alpha$ for two samples of QSOs 
(dots) and BL Lac Objects (open circles). 
QSOs \& BL Lac Objects appear clearly separated.
Model predictions for both classes of objects are discussed in the text:
black body (dot-dashed line), effect of the host galaxy (thin line), change of
accretion rate (thick dashed line), hot spots (large filled squares), synchrotron
emission for BL Lac Objects (vertical lines).}
               \label{fig1}%
     \end{figure*}

\section{BL Lac Objects}
The continuum spectral energy distribution of blazars from radio frequencies
to X and $\gamma$-rays can be explained by a synchrotron emission
plus inverse Compton scattering \citep{sik94}. Variability
can be produced by an intermittent channeling into the jet of the energy produced by 
the central engine.
\citet{spa01} have considered a detailed model where 
crossing of different shells of material, ejected with different velocities,
produce shocks which heat the electrons responsible for the synchrotron emission.
The resulting spectra are compared with multi-band, multi-epoch observations of
3C 279 from radio to $\gamma$ frequencies, showing a good agreement.
In the case of the eight objects of our sample,  B, V, R, I bands
are  sampling variability of the synchrotron component.
This component can be roughly described by a broken power law 
characterized by the break (or peak in $\nu L_{\nu}$)
frequency $\nu_p$ and the asymptotic spectral slopes
$\alpha_1$ end $\alpha_2$ at low and high frequency respectively,
$\alpha_1 > -1$, $\alpha_2 < -1$  \citep{tav98}.
We adopt \citep{vag02} the equivalent representation
\begin{eqnarray}
L_{\nu}={2L_p\over
(\frac{\nu}{\nu_p})^{-\alpha_1}+(\frac{\nu}{\nu_p})^{-\alpha_2}} \equiv \nonumber
\\
\equiv L_p\cdot H(\nu;\alpha_1,\alpha_2,\nu_p)
\end{eqnarray}
for a stationary component of the SED and we add to it
a second (variable) component with the same analytical
form but different peak frequency $\nu_{p'}$ and amplitude $L_{p'}$
to produce spectral changes:
\begin{eqnarray}
L_{\nu}= L_p\cdot H(\nu; \alpha_1, \alpha_2, \nu_p) + \nonumber \\
+ L_{p'}\cdot H(\nu; \alpha_1, \alpha_2, \nu_{p'})
\end{eqnarray}

The addition of the second component mimics the behavior of the synchrotron emission
of model spectra when shell crossing occurs,  producing an increment of emission with
$\nu_{p'} > \nu_p$, due to newly accelerated electrons.
With such a representation we can compute $\alpha$ and $\beta$ as a function of 
$\nu_{p'}/\nu_p$, for different values of $\nu_p$ and given values
of $\alpha_1$, $\alpha_2$ and $L_{p'}/L_p$.  
We adopt typical values of the asymptotic slopes \citep{tav98}
$\alpha_1=- 0.5$, $\alpha_2=-1.75$,
we assigned to $\nu_p$ different values in the range $10^{14}-10^{15}$,
we made $\nu_{p'}/\nu_p$ vary in the range $\approx 1-10$ and we adopted
$L_{p'}/L_p$  corresponding to a magnitude change of 1 mag r.m.s.
The results are shown in Fig. 1, where for $\alpha$ we use the slope of the 
stationary component. The three vertical lines are computed for
$\log\nu_p=14,14.5,14.8$ from left to right respectively. 
The computed lines fall naturally in the region occupied by the
data, namely it is possible to account for the position of the objects
in the $\alpha-\beta$ plane with typical values of
$\alpha_1$, $\alpha_2$, $\nu_{p'}/\nu_p$ and $\nu_p$,
corresponding to the overal SED of the objects considered \citep[see][]{fos98}.

\section{Conclusions}
We show that the spectral variability parameter $\beta$ is a powerful tool to 
discriminate between different models of the variability of AGNs. 
Hot spots on the disk, likely produced by local instabilities, are able
to account for the observed spectral variability of QSOs.

We show that  BL Lacs clearly differ from QSOs in their
$\alpha,\beta$ distribution. A simple model 
representing the variability of a synchrotron component
can account for the observed $\alpha$ and $\beta$ values.

In the framework of wide field variability studies, we stress that 
observations in at least two photometric bands,
repeated on the same field at many epochs, would allow a detailed
test of variability models, extending our knowledge of the emission
processes in AGNs.

\bibliographystyle{aa}
{}

\end{document}